\documentstyle[multicol,aps,prl,epsf,rotate]{revtex}

\date{\today}
\draft
\tighten
\newcommand{\be}{\begin{equation}}
\newcommand{\ee}{\end{equation}}
\newcommand{\bea}{\begin{eqnarray}}
\newcommand{\eea}{\end{eqnarray}}
\begin{document}
\def\sqr#1#2{{\vcenter{\hrule height.3pt
      \hbox{\vrule width.3pt height#2pt  \kern#1pt
         \vrule width.3pt}  \hrule height.3pt}}}
\def\square{\mathchoice{\sqr67\,}{\sqr67\,}\sqr{3}{3.5}\sqr{3}{3.5}}
\def\today{\ifcase\month\or
  January\or February\or March\or April\or May\or June\or July\or
  August\or September\or October\or November\or December\fi
  \space\number\day, \number\year}

\def\Bbb{\bf}
\topmargin=-0.3in


\newcommand{\ww}{\mbox{\tiny $\wedge$}}
\newcommand{\pp}{\partial}

\title{Interaction of a TeV Scale Black Hole with the Quark-Gluon
Plasma at LHC }
\author{Andrew Chamblin$^{1}$, Fred Cooper$^2$,
and Gouranga C. Nayak$^{2}$}
\address{
$^{1}$Department of Physics,
Queen Mary, University of London,
Mile End Road, London E1 4NS \\
$^{2}$T-8, Theoretical Division, Los Alamos National Laboratory,
Los Alamos, NM 87545, USA}
\maketitle

\begin{abstract}
\\
If the fundamental Planck scale is near a TeV, then parton collisions with
high enough center-of-mass energy should produce black holes. The production
rate for such black holes has been extensively studied for the case
of a proton-proton collision at $\sqrt s$ = 14 TeV and for a lead-lead collision
at $\sqrt s$ = 5.5 TeV at LHC. As the parton energy
density is much higher at lead-lead collisions than in pp collisions at LHC,
one natural question is whether the produced black holes will be able to absorb the partons
formed in the lead-lead collisions and eventually `eat' the quark-gluon plasma
formed at LHC.
In this paper, we make a quantitative analysis of this possibility and
find that since the energy density of partons formed in lead-lead collisions
at LHC is about 500 GeV/fm$^3$, the rate of absorption for one of these black holes
is much smaller than the rate of evaporation. Hence, we argue that black holes formed
in such collisions will decay very quickly, and will not absorb very many
nearby partons. More precisely, we show that for the black hole mass to increase via parton absorption
at the LHC the typical energy density of quarks and gluons should be of the
order of $10^{10}$ GeV/fm$^3$. As LHC will not be able to produce such a
high energy density partonic system, the black hole will not be able to
absorb a sufficient number of nearby partons before it decays. The typical
life time of the black hole formed at LHC is found to be a small fraction of a $fm/c$.

\end{abstract}

\begin{multicols} {2}

\section{Introduction}

There is now a huge literature devoted to the possibility that the
scale of quantum gravity
could be as low as a TeV \cite{folks}. If this is true,  then quantum gravity
and perhaps even string theory effects will become relevant in collider
experiments.  For example, there are many discussions on graviton
and radion production and processes related to them at LHC \cite{gravr,gravr1}.
One of the most exciting aspects of this TeV scale gravity
will be the production of black holes in particle accelerators. These
`brane-world' black holes will be our first window into the extra dimensions
of space predicted by string theory, and required by the brane-world
scenarios that provide for a low energy Planck scale \cite{folks}.
While the exact metrics describing black holes in brane-world scenarios are
unknown, considerable work on this issue is underway \cite{large}.
Furthermore, even without the exact metrics it is possible to make
estimates based on crude information. In particular, it is well-understood that
when the mass of the black hole is greater than the Planck scale,
the gravitational
field of the brane can be neglected; furthermore, as long as the size of the
black hole is small compared to the characteristic length scales, then
a brane-world black hole may be regarded, to very good approximation, as simply
a higher-dimensional black hole in flat space.
Using these approximations, in a number of recent
papers people have studied the production of microscopic black holes in
proton-proton (pp) and lead-lead (PbPb) collisions at LHC and cosmic ray events
\cite{ppbf,pp,pp1,pp2,pp3,ag,ppch,ppk,ppu,park,hof,more}, \cite{cham},
\cite{pp4,pp5,pp6}.

The evolution of a black hole after its formation is governed by two
competing effects: 1) absorption of nearby particles by the black
hole and 2) evaporation of the black hole.  Many papers have been written
for the case of primordial black holes within the braneworld
scenarios \cite{bran}. In this paper we will study the case of the black holes
formed at LHC.  In addition to any black holes formed at LHC in pp or
PbPb collisions, there will be many more standard model particles formed, in
particular a huge number of quarks and gluons.
The energy density of the quarks and gluons formed in a pp collision
is not very large and hence the absorption of partons by the black hole
is very small. However, the energy density of the partons formed in the
PbPb collisions at LHC (at $\sqrt s$ = 5.5 TeV per nucleon)
is large. This is because a huge number of
partons are formed in PbPb collisions (equivalent to 208-208 pp collisions)
in a small volume. Therefore one needs to determine if a black hole
can absorb a sufficient amount of partons after its formation in PbPb
collisions to eventually `eat' the quark-gluon plasma. In this
paper we will analyze the evolution of the black hole at LHC by incorporating
the absorption of partons by the black hole and the decay of the
black hole simultaneously.

The paper is organized as follows: In section II we present the calculation
of the black hole mass evolution at LHC.
In section III we present a perturbative QCD calculation to determine the
energy density of the partons formed
in PbPb collisions at LHC. In section IV and V
we present our main results and conclusions respectively.

\section{Evolution of a TeV Scale Black Hole at LHC}

The absorption rate of the black hole is proportional to the surface area
of the black hole and to the energy density of the mass less gluons
produced at LHC \cite{brane}. The effective black hole radius $r_{eff}$
for capturing particles is given by \cite{grey4}:
\bea
r_{eff}~=~ \frac{1}{M_P} \sqrt{\frac{d+3}{d+1}}~
[\frac{2^d \pi^{(d-3)/2} (d+3) \Gamma(\frac{d+3}{2})}{2(d+2)}
\frac{M_{BH}}{M_P}]^{\frac{1}{1+d}}
\eea
where $M_P$ is the TeV scale Planck mass, $M_{BH}$ is the black hole mass
and $d$ is the number of extra dimensions. The effective radius in various
numbers of extra dimensions is given as shown:
\bea
r_{eff}~=~1.6 \frac{1}{M_P} [\frac{M_{BH}}{M_P}]^{\frac{1}{3}}~~~~~{\rm for}~~~~~d=2 \nonumber \\
r_{eff}~=~2.63 \frac{1}{M_P} [\frac{M_{BH}}{M_P}]^{\frac{1}{5}}~~~~~{\rm for}~~~~~d=4  \nonumber \\
r_{eff}~=~3.77 \frac{1}{M_P} [\frac{M_{BH}}{M_P}]^{\frac{1}{8}}~~~~~{\rm for}~~~~~d=7.
\eea

The cross section for the particle absorption by the black hole
is given by:
\be
\sigma = \pi r_{eff}^2.
\ee
Using this absorption cross section the
accretion rate of the black hole mass becomes \cite{brane}:
\be
\frac{dM_{BH}}{d\tau}|_{accr}~=~F \pi r^2_{eff} \epsilon(\tau),
\ee
where
$\epsilon(\tau)$ is the energy density of the mass less gluons produced at
LHC. As gluons are the dominant part of the total parton production at LHC
(more than 80 percent) we will consider the absorption of the
gluons in this paper for simplicity.
The energy density $\epsilon(\tau)$ of the gluons formed in PbPb collisions
at LHC will be calculated in the next section by using perturbative QCD.
For an upper limit on our estimates on the accretion rate
we will take $F~=~1$ in our calculation.
This value might be lower indicating less accretion \cite{brane}. As we are
interested in the maximum accretion we will use $F=1$ in this paper.

Now we consider the evaporation rate of the black hole at LHC. It is known
that the evaporation is much faster for smaller mass black holes
\cite{rad}
and hence the rate of evaporation of the TeV scale
black hole at LHC is much higher. We use the recent calculations
of black hole radiation in the context of TeV scale physics 
\cite{bran,rad,rad1} which yields the evaporation rate:
\bea
&&\frac{dM_{BH}}{d\tau}|_{eva}~=~-\sum
\Gamma_{s} g_{eff}^s \frac{\pi^2}{120}A_{3+d}
T_{BH}^{4+d}~=~\nonumber \\
&&- \sum \Gamma_{s} g_{eff}^s \frac{1}{7680}\frac{(\frac{d+3}{2})^{(d+3)/(d+1)}
(d+1)^{d+3}}
{4^d \pi^{\frac{d+1}{2}} (\frac{d+1}{2})!} \frac{1}{r^2_{eff}}
\label{eva}
\eea
where
\be
T_{BH}=\frac{d+1}{4\pi r_{BH}}
\ee
with $r_{BH}$ being the Schwarzschild radius of the black hole:
\bea
r_{BH}~=~ \frac{1}{M_P} 
[\frac{2^d \pi^{(d-3)/2} \Gamma(\frac{d+3}{2})}{(d+2)}
\frac{M_{BH}}{M_P}]^{\frac{1}{1+d}}
\eea
and the area in the extra dimension is given by:
\be
A_{3+d}~=~\frac{(3+d)\pi^{\frac{d+3}{2}}}{(\frac{d+3}{2})!}r_{BH}^{2+d}.
\ee
As the change in Stefan-Boltzmann constant is very weakly dependent on the
number of extra dimension \cite{rad} we have used the four dimensional
value equals $\frac{\pi^2}{120}$ in the above.
In the equation (\ref{eva})
$g_{eff}$ corresponds to the effective degrees of freedom of the standard
model particles into which the black hole evaporates. The effective number
of degrees of freedom is given by: $g_{eff}=g_{\rm{bos}}~+~\frac{7}{8}~
g_{\rm{ferm}}$. In this paper we will assume that a black hole radiates
mainly to the mass less particles (electrons and positrons, muons and anti-muons,
photons, neutrions, gluons and up, down, strange quarks) for which
$g_{eff}~=~46$ with $g_{eff}$=18 for spin $s=1$ and 28 for $s=\frac{1}{2}$.
$\Gamma_s$ is the dimensionless grey body factor which we take the same
value as in \cite{grey4} in the geometric optics approximations. There has
been recent progress on grey body factors in extra dimensions \cite{greyd}.
$\sum$ in the above equation is the sum of fermions and bosons. For thermodynamics
to be a reasonable approximation one requires $M_{BH} >> M_P$ \cite{grey4},
which will be true for the ratio being $\ge 3$.

The net change of the black hole mass is therefore:
\be
\frac{dM_{BH}(\tau)}{d\tau}~=~ \frac{dM_{BH}(\tau)}{d\tau}|_{accr}~+~
\frac{dM_{BH}(\tau)}{d\tau}|_{eva}
\ee
which gives:
\be
\frac{dM_{BH}(\tau)}{d\tau}~=~
 \pi r^2_{eff}(\tau) \epsilon(\tau)
- \frac{\Gamma_{s} g_{eff}^s} {7680} \frac{f_d}{r^2_{eff}}
\label{net1}
\ee
where $f_d= \frac{(\frac{d+3}{2})^{(d+3)/(d+1)} (d+1)^{d+3}}
{4^d \pi^{\frac{d+1}{2}} (\frac{d+1}{2})!}$. We get $f_2$=9.45,
$f_4$=30 and $f_7$=210.
By using the relation between $r_{eff}$ and the black hole mass $M_{BH}$ we obtain,
from the above equation:
\be
\frac{dM_{BH}(\tau)}{d\tau}~=~
F_d(\tau)~M_{BH}^{\frac{2}{d+1}}(\tau)
~-~\frac{B_d}{M_{BH}^{\frac{2}{d+1}}(\tau)}.
\label{net}
\ee
This equation has to be solved numerically once we know the gluon
energy density $\epsilon(\tau)$ at LHC. In the above equation
\bea
&&F_d~=~\pi \epsilon(\tau)~ C_d \nonumber \\
&&B_d~=~\frac{\sum \Gamma_s
g_{eff}^s f_d}{ 7680 C_d} ~~~~~~~~{\rm with}~~~~~~~~~\nonumber \\
&&C_d=\frac{d+3}{(d+1)M_P^2}
~[\frac{2^d \pi^{(d-3)/2} (d+3)
\Gamma(\frac{d+3}{2})}{2(d+2)M_P}]^{\frac{2}{1+d}}.
\eea
We now just need to determine the gluon energy density at LHC by using perturbative QCD methods.

\section{Gluon energy density at LHC}

Now let us determine the energy density of the mass less gluons formed
in PbPb collisions at LHC. We use perturbative QCD methods
to determine the gluon energy density produced at LHC.
In the lowest order of pQCD calculations the cross section
for gluon-minijet production in pp collisions
is given by \cite{kaj,naykk,nayk}:
\bea
&&\sigma_{jet} = ~\int dp_T dy_1 dy_2 {{2 \pi p_T} \over {\hat{s}}}
\sum_{ijkl} x_1~ f_{i/A}(x_1, p_T^2)~ \nonumber \\
&&x_2~ f_{j/A}(x_2, p_T^2)~
\hat{\sigma}_{ij \rightarrow kl}(\hat{s}, \hat{t}, \hat{u}).
\label{jet}
\eea
The jet cross section can be related to the total number of minijets
formed in PbPb collisions at LHC:
\bea
&&N_{jet} = T(0)~\int dp_T dy_1 dy_2 {{2 \pi p_T} \over {\hat{s}}}
\sum_{ijkl} x_1~ f_{i/A}(x_1, p_T^2)~ \nonumber \\
&&x_2~ f_{j/A}(x_2, p_T^2)~
\hat{\sigma}_{ij \rightarrow kl}(\hat{s}, \hat{t}, \hat{u}),
\label{number}
\eea
where $T(0)= 9A^2/{8\pi R_A^2}$ is the nuclear geometrical factor for
head-on PbPb collisions (for a nucleus with a sharp surface).
Here $R_A=1.2 A^{1/3}$ fm is the nuclear radius.
$\hat{\sigma}_{ij \rightarrow kl}$ denotes the elementary pQCD parton cross
sections, which are given by:
\begin{equation}
\hat{\sigma}_{gq \rightarrow gq} = {{\alpha_s^2} \over
{ \hat{s}}} ({\hat{s}^2+\hat{u}^2}) [
{{1} \over {\hat{t}^2}}
- {{4} \over {9\hat{s}\hat{u}}}],
\nonumber
\end{equation}
\begin{equation}
\hat{\sigma}_{gg \rightarrow gg} = {{9 \alpha_s^2} \over
{2 \hat{s}}} [ 3 - {{\hat{u}\hat{t}} \over {\hat{s}^2}}
 - {{\hat{u}\hat{s}} \over {\hat{t}^2}}
 - {{\hat{s}\hat{t}} \over {\hat{u}^2}}].
\nonumber
\end{equation}
As gluons are the dominant part of the parton production at LHC we have
considered the above two partonic type collisions.
The kinematic relations are given by:
\begin{equation}
\hat{s}=x_1 x_2 s = 4 p_T^2 ~{\rm cosh}^2
\left ( {{y_1-y_2} \over {2}} \right ).
\nonumber
\end{equation}
The momentum rapidities $y_1$, $y_2$ and the momentum
fractions $x_1$, $x_2$ are related by,
\begin{equation}
x_1=p_T~(e^{y_1}+e^{y_2})/{\sqrt{s}}, \hspace{0.5cm}
x_2=p_T~(e^{-y_1}+e^{-y_2})/{\sqrt{s}}.
\nonumber
\end{equation}
We use the saturation argument to fix the minimum momentum scale as
$p_0$ above which the pQCD is applicable to be 2 GeV at LHC \cite{muller}.
The partonic structure functions $f_{i,j/A}$ used in Eq.\ (\ref{jet}) are
GRV98 \cite{grv98} for the proton structure function and EKS98 \cite{eks98}
for nuclear shadowing effects.

The total transverse energy $<E_T^{tot}>$ produced in nuclear collisions
is given by:
\bea
&&<E_T^{jet}> = T(0) \int dp_T p_T dy_1 dy_2 {{2 \pi p_T}
 \over {\hat{s}}} \sum_{ijkl} x_1 ~f_{i/A}(x_1, p_T^2) ~ \nonumber \\
&&x_2 ~f_{j/A}(x_2, p_T^2)
 ~\hat{\sigma}_{ij \rightarrow kl}(\hat{s}, \hat{t}, \hat{u}).
\label{energy}
\eea
The above pQCD formula has the information in the momentum space via
$p_T$ and $y$ (the momentum rapidity) distribution. In order to obtain
energy density we need to supply additional information about coordinate
distribution. For an expanding system the minijet number distribution
eq. (\ref{number}) can be related to the phase-space distribution
function via \cite{ref:c-f}:
\be
\frac{d^3 N^{jet}}{\pi dy dp_T^2} = g_C
\int f(x, p) ~p^{\mu} d\sigma_{\mu}
\ee
where $g_C=16$ is the product of spin and color degrees of freedom,
\bea
d\sigma^{\mu}=\pi R_A^2 \tau d\eta (\cosh \eta,0,0, \sinh \eta),
\eea
and
\bea
p^{\mu}= (p_T \cosh y, p_T \cos \phi, p_T \sin \phi, p_T \sinh y),
\eea
for an 1+1 dimension expanding system.
Using the above relations we get at the initial time $\tau_0$ (=1/$p_0$):
\be
\frac{d N^{jet}}{\pi dy dp_T^2} = g_C \pi R_A^2 \tau_0 \int d\eta
~p_T \cosh(\eta-y)~ f(p_T, \eta, y, \tau_0),
\label{nb}
\ee
where we assume a transverse isotropy at the early stage.
We take a boost non-invariant distribution function in the following.
We assume a Gaussian coordinate rapidity $\eta$
(=$[\frac{1}2{}\rm{ln}\frac{t+z}{t-z}]$) distribution of the form \cite{nayk}
\be
f(p_T, \eta, y, \tau_0) = f(p_T,y)
e^{-\frac{(\eta-y)^2}{\sigma^2(p_T)}},
\label{inds1}
\ee
where $f(p_T,y)$ is obtained from
$\frac{d N^{jet}}{dy dp_T}$ by using pQCD from eq. (\ref{number}).
Using the above form in eq. (\ref{nb}) we get:
\bea
\frac{d N^{jet}}{dy dp_T}&& = g_C 2 \pi^2 R_A^2 \tau_0 ~p_T^2 f(p_T,y)
\int_{-\infty}^{\infty} d\eta \cosh (\eta-y) \times \nonumber \\
&& e^{-\frac{(\eta-y)^2}{\sigma^2(p_T)}}
\label{nbt1}
\eea
which gives:
\be
\frac{d N^{jet}}{dy dp_T} = g_C
2 \pi^2 \sqrt{\pi} R_A^2 \tau_0 p_T^2
f(p_T,y)\sigma(p_T) e^{\sigma^2(p_T)/4}.
\ee
~From the above equation we get:
\be
f(p_T,y) =
\frac{\frac{d N^{jet}}{dy dp_T}}{ g_C 2 \pi^2
\sqrt{\pi} R_A^2 \tau_0 p_T^2
\sigma(p_T) e^{\sigma^2(p_T)/4}}
\ee
which gives a boost non-invariant initial phase-space gluon
distribution function:
\be
f(p_T, \eta, y, \tau_0) =
\frac{\frac{d N^{jet}}{dy dp_T}}{ g_C 2 \pi^2 \sqrt{\pi} R_A^2
\tau_0 p_T^2 }
\frac{e^{-\frac{(\eta-y)^2}{\sigma^2(p_T)}}}{\sigma(p_T)
e^{\sigma^2(p_T)/4}}.
\ee
Using the above boost non-invariant gluon distribution function
the initial minijet energy density is given by:
\bea
&&\epsilon(\tau_0,\eta)= g_C \int d\Gamma (p^{\mu}u_{\mu})^2
f(p_T, \eta, y,\tau_0) = g_C \int d^2p_T p_T^2 \nonumber \\
&& \int dy \cosh^2 (\eta-y) f(p_T,\eta, y,
\tau_0) = 2 \pi \int dp_T p_T \nonumber \\
&& \int dy \frac{\frac{d N^{jet}}{dy dp_T}}{ 2 \pi^2 \sqrt{\pi} R_A^2 \tau_0 }
\frac{e^{-\frac{(\eta-y)^2}{\sigma^2(p_T)}}}{\sigma(p_T)
e^{\sigma^2(p_T)/4}} \cosh^2(\eta-y).
\label{ednb}
\eea

Before we calculate the initial gluonic energy density at LHC
we would like to find out
the values of the unknown parameter $\sigma^2$ appearing in the above
equation. We will fix this unknown parameter by equating the first $p_T$
moment of the distribution function with the pQCD predicted $E_T$
distribution as given by eq. (\ref{energy}). We have
\bea
&&\int dp_T \int dy \frac{dE_T^{jet}}{dy dp_T}=
2g_C\pi^2 R_A^2 \tau_0 \int
dp_T \int dy ~p_T^3 f(p_T,y) \nonumber \\
&&= \int dp_T \int dy \frac{p_T
\frac{d N^{jet}}{dy dp_T}}{ \sqrt{\pi} \sigma(p_T) e^{\sigma^2(p_T)/4}}.
\label{nbn}
\eea
We determine the value of $\sigma^2$ by using the pQCD values of
$ \frac{dE_T^{jet}}{dy dp_T}$
from eq. (\ref{energy}) in the left hand side of the above equations.
Assuming $\sigma^2$ independent of $p_T$ we find from the above
equation $\sigma^2$ = 0.28 at LHC. Using this values of the $\sigma^2$ we find
from eq. (\ref{ednb}): $\epsilon_0$=517 GeV/fm$^3$ at
LHC in the central rapidity region.

\section{Results and Discussions}

In this section we will present the results for the evolution of the
black hole mass at LHC by taking both accretion and radiation into
account. The accretion rate depends on the energy density of the gluons
(see eq. (\ref{net})) which is a function of time. The energy density
of the gluons at LHC decreases as a function of time. This is because
of the fast expansion of the system, which acquires more volume. In our
numerical calculation we will assume a maximum energy density at
all time which is the energy density obtained at initial collision,
as described in the last section. Thus our results will overestimate the effect
of accretion and thus underestimate the critical energy density. However 
the expansion time is at a fm/c scale which is slow compared to the 
evaporation time scale at LHC. 
We solve the eq. (\ref{net}) numerically with the gluon energy density
obtained in the last section. The time evolution of a black hole
of mass 3 TeV is shown in Fig.1. The Planck mass is choosen to be 1 TeV.
Different curves are for different numbers of extra dimensions. Solid, dotted
and dashed lines correspond to d=2, 4 and 7 respectively. It can
be seen that the decay of a 3 TeV mass 
black hole is almost instantaneous at LHC even
if one takes into account the absorption of partons with energy density
of 517 GeV/fm$^3$. The black hole of mass 3 TeV (with Planck mass equals to 
1 TeV) decays within a fraction of a $fm/c$. The partons in the QGP phase 
hadronize in a hadronization time scale which is more than a $fm/c$. 
Hence a 3 TeV mass black hole completely evaporates 
before any of the hadrons are formed at LHC. 

Now let us study the evolution of a heavier black hole, say 5 TeV,
at LHC. As stated earlier a heavier mass black hole will evaporate 
slower. This is because the temperature of the black hole is inversely
proportional to its mass. The evolution of a 5 TeV mass black hole at LHC
is shown in Fig. 2. We have used a Planck mass equals to 1 TeV. In comparison
to Fig. 1 the decay time of the 5 TeV black hole mass is slightly enhanced
when compared to the case when the mass of the black hole is
3 TeV, but it is still instantaneous in comparison
to hadronization scale which is more than a $fm/c$. This suggests that
a typical black hole formed at LHC evaporates instantaneously and
can not eat sufficient matter. 

In Fig. 3 we have shown the results of the time evolution of the 5 TeV
mass black hole but with Planck mass equals to 2 TeV. As can be seen
from the figure the evaporation is much faster than the earlier two
cases, considered in Fig. 1 and 2. This is because as the Planck mass is
increased the Schwarzschild radius is decreased. Due to the decrease
in the Schwarzschild radius the effective cross section for absorption
of matter by the black hole is decreased. Hence the evolution of the
black hole mass decreases much faster as we increase the Planck mass
which can be seen from Fig. 3. 

In the following we will present estimates of the minimum energy density
of the matter 
we would need in order for a TeV scale black hole to swallow quark-gluon
matter at LHC. In order for a black hole to increase its mass, one must have:
\be
\frac{dM_{BH}}{d\tau}~>~ 0.
\ee
Hence from eq. (\ref{net1}) we get the minimum required energy density 
($\epsilon_{min}$) of the partonic matter to be:
\be
\epsilon_{min}~=~
\frac{\sum \Gamma_s g_{eff}^s f_d}{7680 \pi r^4_{eff}}.
\label{emin}
\ee

From the above equation we get for $M_P$ = 1 TeV: \\

1) Case with the number of extra dimensions d=2:

\bea
\epsilon_{min}~=~4.2 \times 10^{10}~~ GeV/fm^3~~~~{\rm for}~~~~M_{BH}= 3 ~TeV \nonumber \\
\epsilon_{min}~=~2.1 \times 10^{10}~~ GeV/fm^3~~~~{\rm for}~~~~M_{BH}= 5 ~TeV, \nonumber
\eea

2) Case with the number of extra dimensions d=4:

\bea
\epsilon_{min}~=~3.2 \times 10^{10}~~ GeV/fm^3~~~~{\rm for}~~~~M_{BH}= 3 ~TeV \nonumber \\
\epsilon_{min}~=~2.2 \times 10^{10}~~ GeV/fm^3~~~~{\rm for}~~~~M_{BH}= 5 ~TeV, \nonumber
\eea

3) Case with the number of extra dimensions d=7:

\bea
\epsilon_{min}~=~7.5 \times 10^{10}~~ GeV/fm^3~~~~{\rm for}~~~~M_{BH}= 3 ~TeV \nonumber \\
\epsilon_{min}~=~5.8 \times 10^{10}~~ GeV/fm^3~~~~{\rm for}~~~~M_{BH}= 5 ~TeV. \nonumber
\eea
It can be seen that the minimum energy density required for different
values of $M_{BH}$ and for different values of number of extra dimension
is much higher than the energy density obtained at LHC.
Note that these numbers are obtained for $M_P$= 1 TeV. For higher values of
$M_P$ these numbers will be further increased. Creation of
such astronomically high values of energy density is beyond the reach
of any accelerator experiment. This means that black holes created in high
energy hadronic colliders at LHC will not swallow the matter nearby.

\section{Conclusions}

If the fundamental Planck scale is near a TeV, then parton collisions with
high enough center-of-mass energy should produce black holes. The production
rate for such black holes has been extensively studied for the case
of a proton-proton collision at $\sqrt s$ = 14 TeV and for a lead-lead collision
at $\sqrt s$ = 5.5 TeV at LHC. As the parton energy
density is much higher at lead-lead collisions than in pp collisions at LHC,
it is necessary to check
if the produced black hole will be able to absorb the partons
formed in the lead-lead collisions and eventually consume much of 
the quark-gluon plasma formed at LHC.
In this paper, we have made a quantitative analysis of this possibility and
found that as the energy density of partons formed in lead-lead collisions
at LHC is about 500 GeV/fm$^3$, the rate of absorption of these black holes
is much smaller than the rate of evaporation. Hence black holes formed
in such collisions evaporate much too quickly to absorb the
partons nearby. For the black hole mass to increase via parton absorption
at the LHC the typical energy density of quarks and gluons should be of the
order of $10^{10}$ GeV/fm$^3$. As LHC will not be able to produce such a
high energy density partonic system, the black hole will not be able to
absorb sufficient nearby partons before it decays. The typical
life time of the black hole formed at LHC is found to be a small
fraction of a $fm/c$.

\acknowledgments
We thank Haim Goldberg for careful reading of the manuscript and for
useful suggestions. We thank Emil Mottola and Stavros Mouslopoulos for 
discussions. AC was supported by a Director's Funded Fellowship at Los Alamos
National Lab,
where this research was supported by the Department of Energy, under
contract W-7405-ENG-36, and is currently a PPARC Advanced Research Fellow.

\end{multicols}

\newpage

\vspace{2.5cm}

\parbox{15cm}{
\parbox[t]{7cm}
{\begin{center}
\mbox{\epsfxsize=6.5cm\epsfysize=4cm\epsfbox{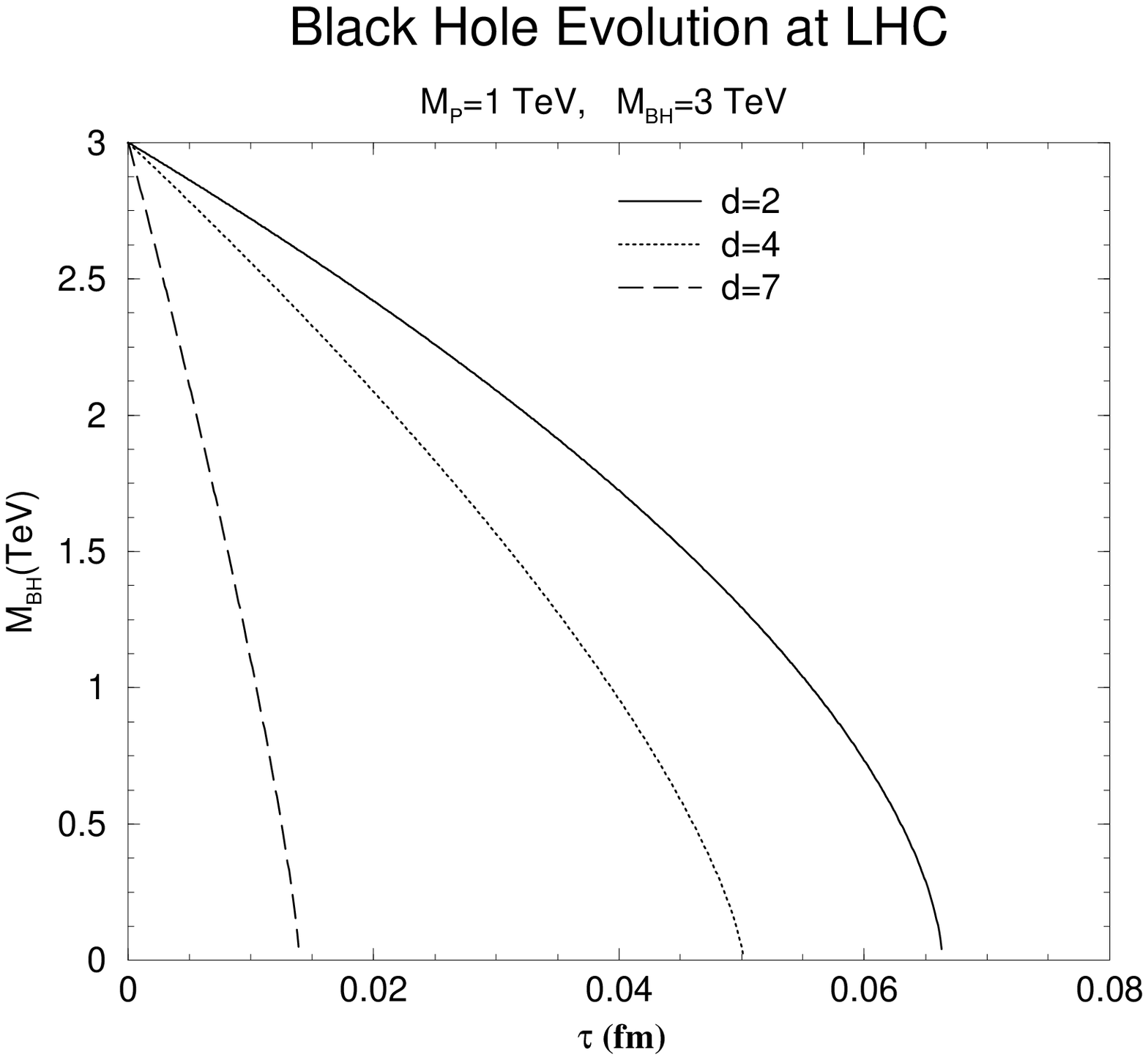}}
\end{center}}
\hspace{.5cm}\parbox[t]{7cm}
{\begin{center}
\mbox{\epsfxsize=6.5cm\epsfysize=4cm\epsfbox{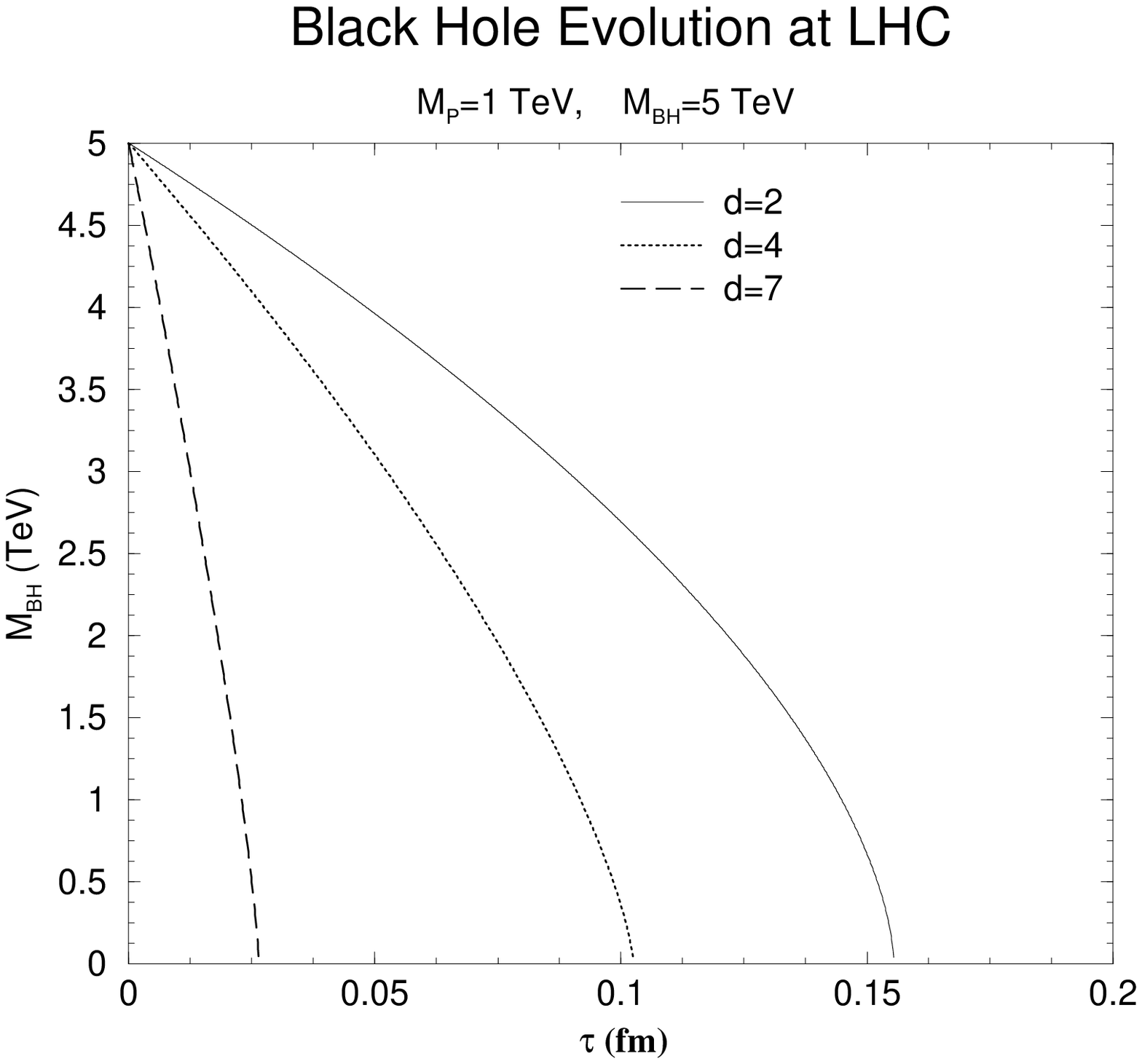}}
\end{center}}
\parbox[t]{7cm}
{{\small FIG. 1: Evolution of black hole of mass equals to
3 TeV at LHC. The Planck mass equals to 1 TeV.
}}
\hspace{0.8cm}\parbox[t]{7cm}
{{\small FIG. 2: Evolution of black hole of mass equals to 5 TeV
at LHC. The Planck mass equals to 1 TeV.
}}}

\vspace{2.5cm}

\parbox{15cm}{
\parbox[t]{7cm}
{\begin{center}
\mbox{\epsfxsize=6.5cm\epsfysize=4cm\epsfbox{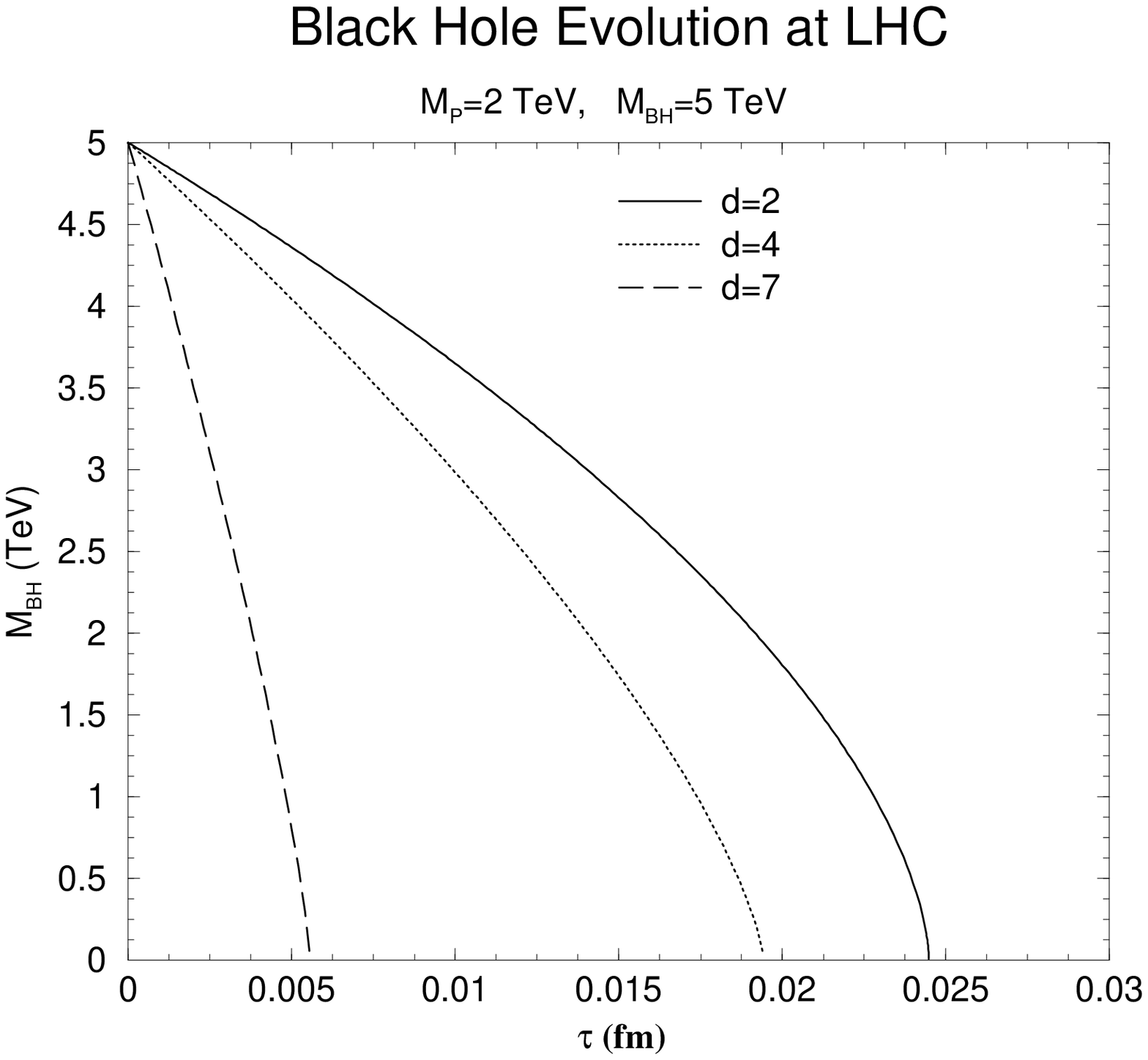}}
\end{center}}
{\begin{center}
\end{center}}
\parbox[t]{7cm}{{\small FIG. 3: Evolution of black hole of mass equals to
5 TeV at LHC. The Planck mass equals to 2 TeV.
}}}


\begin{references}

\bibitem{folks}
N. Arkani-Hamed, S. Dimopoulos and G. Dvali,
Phys. Lett. {\bf B429}, 263 (1998);
I. Antoniadis, N. Arkani-Hamed, S. Dimopoulos and G. Dvali,
Phys. Lett. {\bf B436}, 257 (1998);
L.~Randall and R.~Sundrum, Phys. Rev. Lett. {\bf 83}, 3370 (1999);
L.~Randall and R.~Sundrum, Phys. Rev. Lett. {\bf 83}, 4690 (1999).

\bibitem{gravr} G. F. Giudice, R. Rattazzi and J. D. Wells, Nucl. Phys. B
544 (1999) 3; L. Vacavant and I. Hinchliffe, 
J. Phys. G27 (2001) 1839; G. C. Nayak, hep-ph/0211395; 
Y. A. Kubyshin, hep-ph/0111027; 
S.~B.~Bae et al., Phys.\ Lett.\ B {\bf 487}, 299 (2000);
S. C. Park, H. S. Song and J. Song, Phys. Rev. D63 (2001) 077701;
K.~Cheung, Phys.\ Rev.\ D {\bf 63}, 056007 (2001).

\bibitem{gravr1} 
W.~D.~Goldberger and M.~B.~Wise,
Phys.\ Rev.\ Lett.\  {\bf 83}, 4922 (1999);
Phys.\ Lett.\ B {\bf 475}, 275 (2000);
Phys.\ Rev.\ D {\bf 60}, 107505 (1999);
C.~Cs\'aki, M.~Graesser, L.~Randall and J.~Terning,
Phys.\ Rev.\ D {\bf 62}, 045015 (2000);
C.~Cs\'aki, M.~L.~Graesser and G.~D.~Kribs,
Phys.\ Rev.\ D {\bf 63}, 065002 (2001);
Phys.\ Rev.\ D {\bf 62}, 067505 (2000);
C.~Csaki, J.~Erlich and J.~Terning,
Phys.\ Rev.\ D {\bf 66}, 064021 (2002);
C.~Csaki, J.~Erlich, T.~J.~Hollowood and Y.~Shirman,
Nucl.\ Phys.\ B {\bf 581}, 309 (2000);
E.~A.~Mirabelli, M.~Perelstein and M.~E.~Peskin,
Phys.\ Rev.\ Lett.\  {\bf 82}, 2236 (1999);
D.~Dominici, B.~Grzadkowski, J.~F.~Gunion and M.~Toharia,
arXiv:hep-ph/0206192;
T.~Han, G.~D.~Kribs and B.~McElrath,
Phys.\ Rev.\ D {\bf 64}, 076003 (2001);
M.~Chaichian, A.~Datta, K.~Huitu and Z.~h.~Yu,
Phys.\ Lett.\ B {\bf 524}, 161 (2002);
J.~L.~Hewett and T.~G.~Rizzo, 
{\tt hep-ph/0202155};
For a review see G.~D.~Kribs,
{\tt hep-ph/0110242}.

\bibitem{large} A. Chamblin, S. W. Hawking and H. S. Reall,
Phys. Rev. {\bf D61}, 065007(2000);
R. Emparan, G. T. Horowitz and R. C. Myers, JHEP {\bf 0001},
07(2000); N. Dadhich, R. Maartens, P. Papadopoulos and
V. Rezania, Phys. Lett. B487 (2000) 1; A. Chamblin, H. Reall, H. Shinkai and
T. Shiromizu, Phys. Rev. {\bf D63}, 064015 (2001); P. Kanti and K. Tamvakis,
Phys. Rev. {\bf D65}, 084010 (2002); C. Germani and R. Maartens,
Phys. Rev. {\bf D64}, 124010 (2001); I. Giannakis and H. Ren,
Phys. Rev. {\bf D63}, 125017 (2001); R. Casadio and L. Mazzacurati,
gr-qc/0205129; P. Kanti, I. Olasagasti and K. Tamvakis, Phys. Rev. D66
(2002) 104026.

\bibitem{ppbf} T. Banks and W. Fischler, hep-th/9906038.

\bibitem{pp} S.~Dimopoulos and G.~Landsberg,
Phys.\ Rev.\ Lett.\  {\bf 87}, 161602 (2001).


\bibitem{pp1} S.~B.~Giddings and S.~Thomas,
Phys.\ Rev.\ D {\bf 65}, 056010 (2002).

\bibitem{pp2} S.~B.~Giddings,
in {\it Proc. of the APS/DPF/DPB Summer Study on the Future of Particle Physics
(Snowmass 2001) } ed. R.~Davidson and C.~Quigg, hep-ph/0110127.

\bibitem{pp3} D.~M.~Eardley and S.~B.~Giddings, Phys. Rev. D66 (2002) 044011.


\bibitem{ag}  L. Anchordoqui and H. Goldberg, Phys. Rev. {\bf D65} 047502,
2002.

\bibitem{ppch}  R. Casadio and B. Harms, Int. J. Mod. Phys. A17 (2002) 4635.

\bibitem{ppk}  K. Cheung, Phys. Rev. D66 (2002) 036007; K. Cheung, Phys.
Rev. Lett. 88 (2002) 221602; 
K. Cheung and Chung-Hsien Chou, Phys. Rev. D66 (2002) 036008.

\bibitem{ppu} Y. Uehara, Mod. Phys. Lett. A17 (2002) 1551.

\bibitem{park} Seong Chan Park and H.S.Song, hep-ph/0111069.

\bibitem{hof} M. Bleicher, S. Hofmann, S. Hossenfelder, H. Stoecker, 
Phys. Lett. B548 (2002) 73; S. Hossenfelder, S. Hofmann, M. Bleicher, 
H. Stoecker, Phys. Rev. D66 (2002) 101502. 

\bibitem{more} I. Mocioiu, Y. Nara and I. Sarcevic, hep-ph/0301073;
V. Frolov and D. Stojkovic, gr-qc/0301016; gr-qc/0211055; Phys. Rev. D66
(2002) 084002; D. Ida and S. C. Park, hep-th/0212108; B. Kol, hep-ph/0207037;
T.~Han, G.~D.~Kribs and B.~McElrath, hep-ph/0207003;

\bibitem{cham} A. Chamblin and G. C. Nayak, Phys. Rev. D66 (2002) 091901.


\bibitem{pp4} L. A. Anchordoqui, J. L. Feng, H. Goldberg and A. D. Shapere,
Phys. Rev. D66 (2002) 103002; J.~L.~Feng and A.~D.~Shapere, 
Phys. Rev. Lett. 88 (2002) 021303; 
L. A. Anchordoqui, T. Paul, S. Reucroft and J. Swain,
hep-ph/0206072.

\bibitem{pp5} R.~Emparan, M.~Masip and R.~Rattazzi, Phys. Rev. D65 (2002)
064023.

\bibitem{pp6} A.~Ringwald and H.~Tu,
Phys.\ Lett.\ B {\bf 525}, 135 (2002).

\bibitem{bran} R. Guedens, D. Clancy and A. R. Liddle, Phys. Rev. D66
(2002) 043513 and references therein.

\bibitem{brane} R. Guedens, D. Clancy and A. R. Liddle, Phys. Rev. D66
(2002) 083509 and references therein.

\bibitem{grey4} L. Anchordoqui and H. Goldberg, hep-ph/0209337 

\bibitem{rad} R. Emparan, G. T. Horowitz and R. C. Myers, Phys. Rev.
Lett. 85 (2000) 499.
\bibitem{rad1} M. Fairbairn and V. V. Elewyck, hep-ph/0206257.

\bibitem{greyd} P. Kanti, J. March-Russell, hep-ph/0212199. 

\bibitem{kaj} K. J. Eskola, K. Kajantie and J. Lindfors, Nucl. Phys. B323
(1989) 37.

\bibitem{naykk} G. C. Nayak {\it et al.}, Nucl. Phys. A687 (2001) 457;
R. S. Bhalerao and G. C. Nayak, Phys. Rev. C61 (2000) 054907.

\bibitem{nayk} F. Cooper, E. Mottola and G. C. Nayak, hep-ph/0210391,
Phys. Lett. B (in press).


\bibitem{muller} A. H. Mueller, Nucl. Phys. B572 (2000) 227;
K. J. Eskola, {\it et al.} Nucl. Phys. B570 (2000) 379.

\bibitem{grv98} M. Glueck, E. Reya and A. Vogt, Eur. Phys. J. C5 (1998) 461.

\bibitem{eks98} K. J. Eskola, {\it et al.}, Nucl. Phys. B535 (1998) 351;
Eur. Phys. J. C9 (1999) 61.

\bibitem{ref:c-f} F. Cooper and G. Frye Phys. Rev. D {\bf 10}, 186 (1974).

\end{references}
\end{document}